\documentclass[aps,prx,twocolumn,
			   groupedaddress,superscriptaddress,
			   amsfonts,amssymb,amsmath,
			   citeautoscript,longbibliography,
			   a4paper
			   ]{revtex4-2}

\usepackage[utf8]{inputenc}
\usepackage[english]{babel}

\usepackage{microtype} %For better kerning and symbol-stretching
\usepackage{xspace} %For the \xspace command

\usepackage{txfonts}  %Times Roman fonts
\usepackage{txfontsb} %Addition for txfonts, including old style numerals and greek

\usepackage{bm} %Bold math symbols with \bm{} (Greek and other symbols)

\usepackage{xcolor}
\usepackage[]{graphicx} % "demo" option to disable rendering for speed
\graphicspath{{figs/}}

\usepackage[]{booktabs}
\usepackage{array}
\usepackage{layouts}
\usepackage{multirow}

\usepackage{enumerate}
\usepackage[inline]{enumitem}

\usepackage{hyperref}
\hypersetup{colorlinks,
    linkcolor=black,
    citecolor=black,
    urlcolor=black
}

\usepackage[capitalize,nameinlink]{cleveref}

\crefname{subequations}{Eqs.}{Eqs.} %Specific changes to allow for Eqs.-wording when referring to a set of subequations. Label of subequations must include [subequations] as an option.
\Crefname{subequations}{Eqs.}{Eqs.}
\crefformat{subequations}{#2Eqs.~(#1)#3}
\Crefformat{subequations}{#2Eqs.~(#1)#3}
\crefname{page}{p.}{p.} %Changing from 'page' to 'p.'

\usepackage{placeins}

\usepackage{siunitx}
\DeclareSIUnit[number-unit-product = ]\percent{\char`\%} % remove spacing for \percent

\usepackage[centering,hmargin=16mm,tmargin=30mm,bmargin=26mm]{geometry}

\usepackage{soul}

\usepackage{textcomp} % for \textrightarrow
\usepackage{xifthen}
\usepackage{etoolbox}
\newboolean{togglecomments}
\newboolean{togglechanges}

\setboolean{togglecomments}{true}
\setboolean{togglechanges}{false}

\newcommand{\textblacksquare}{$\blacksquare$}
\newcommand{\todo}[1]{\ifbool{togglecomments}%
	{\textcolor{green!60!black}{\small\textsf{{}\textsuperscript{\textsc{\textsf{todo}}}}[#1]}} % if true, show comments
	{}}     % if false, do nothing
\newcommand{\comment}[2]{\ifbool{togglecomments}%
		{\textcolor{blue!70!black}{\small\sf\textsuperscript{\textsc{\textsf{#1}}}[#2]}} % if true, show comments
		{}}     % if false, do nothing
\newcommand{\swap}[2]{\ifbool{togglechanges}
	{#2}  % revisions-only version
	{\textcolor{red!70!black}{[\ignorespaces#1]}\textrightarrow{}\textcolor{green!50!black}{[\ignorespaces#2]}}}
\newcommand{\remove}[1]{\ifbool{togglechanges}
	{}    % revisions-only version
	{\textcolor{red!70!black}{\ignorespaces#1}}}
\newcommand{\inset}[1]{\ifbool{togglechanges}
	{#1}  % revisions-only version
	{\textcolor{green!50!black}{#1}}}
\newcommand{\citeremind}[1]{%
	[\textcolor{blue!75!black!80!yellow}{\textblacksquare%
		\ifthenelse{\isempty{#1}}{}{\textsuperscript{\tiny\textsf{#1}}}%
	}]\xspace}

\newcommand{\appropto}{\mathrel{\vcenter{
			\offinterlineskip\halign{\hfil$##$\cr
				\propto\cr\noalign{\kern.2pt}\sim\cr\noalign{\kern-2.5pt}}}}}

\makeatletter
\newcommand{\raisemath}[1]{\mathpalette{\raisem@th{#1}}}
\newcommand{\raisem@th}[3]{\raisebox{#1}{$#2#3$}}
\makeatother

\renewcommand{\paragraph}[1]{\vskip 1ex\noindent\textbf{#1.}~}

\usepackage[eulergreek]{sansmath}
\makeatletter
\renewcommand\@make@capt@title[2]{%
    \@ifx@empty\float@link{\@firstofone}{\expandafter\href\expandafter{\float@link}}%
    \sisetup{math-sf=\textsf}%
    \sansmath\sffamily\textbf{#1\@caption@fignum@sep}#2 % does not work with the newtx* packages unfortunately
}%

\makeatother

\newcommand{\mitphysicsaffil}{\footnotesize Department of Physics, Massachusetts Institute of Technology, Cambridge, Massachusetts 02139, USA}
\newcommand{\miteecsaffil}{\footnotesize Department of Electrical Engineering and Computer Science, Massachusetts Institute of Technology, Cambridge, Massachusetts 02139, USA}
\newcommand{\stanfordcsaffil}{\footnotesize Department of Computer Science, Stanford University, Stanford, CA 94305, USA}

\begin{document}

%-----AUTHORS AND AFFILIATIONS-----
\author{Andrew~Ma}
\affiliation{\miteecsaffil}
\author{Owen~Dugan}
\affiliation{\mitphysicsaffil}\affiliation{\stanfordcsaffil}
\author{Marin~Solja\v{c}i\'{c}}
\email{soljacic@mit.edu}
\affiliation{\mitphysicsaffil}

\title{Predicting band gap from chemical composition:\\A simple learned model for a material property with atypical statistics}

\begin{abstract}
    \noindent In solid-state materials science, substantial efforts have been devoted to the calculation and modeling of the electronic band gap.
    While a wide range of ab initio methods and machine learning algorithms have been created that can predict this quantity, the development of new computational approaches for studying the band gap remains an active area of research.
    Here we introduce a simple machine learning model for predicting the band gap using only the chemical composition of the crystalline material.
    To motivate the form of the model, we first analyze the empirical distribution of the band gap, which sheds new light on its atypical statistics.
    Specifically, our analysis enables us to frame band gap prediction as a task of modeling a mixed random variable, and we design our model accordingly.
    Our model formulation incorporates thematic ideas from chemical heuristic models for other material properties in a manner that is suited towards the band gap modeling task.  
    The model has exactly one parameter corresponding to each element, which is fit using data.
    To predict the band gap for a given material, the model computes a weighted average of the parameters associated with its constituent elements and then takes the maximum of this quantity and zero.
    The model provides heuristic chemical interpretability by intuitively capturing the associations between the band gap and individual chemical elements.
    
\end{abstract}

\maketitle

\section{Introduction}

One of the most fundamental properties of a crystalline material is its band gap.
From a scientific point of view, the band gap is essential for characterizing bulk electronic properties -- for example, for a typical material it indicates whether we expect it to be a metal, semiconductor, or insulator.
From a technological point of view, identifying or designing (e.g., via doping) materials with suitable band gaps is crucial for device applications.
Given the importance of this quantity, there has long been substantial research -- both experimental and computational -- devoted to understanding, determining, and engineering the band gap of crystalline materials. 

In computational materials science, the traditional approach to determining the band gap is through ab initio calculations utilizing density functional theory (DFT)~\cite{perdew1985density,sham1985density,tran2009accurate,chan2010efficient,bagayoko2014understanding,xiao2011accurate}.
With the advent of modern computational resources, there have been many works involving the high-throughput DFT calculation of band gaps as well as the benchmarking of various DFT functionals for band gap determination~\cite{garza2016predicting,borlido2019large,kim2020band,jain2013commentary,choudhary2020joint,curtarolo2012aflow,wing2021band,liu2024high}.
While relatively quick calculations of the band gap are possible using inexpensive DFT functionals, such calculations tend to have significant error.
More computationally expensive ab initio calculations are often required to obtain accurate results.
While traditional ab initio methods remain an integral part of computational materials science, the use of machine learning (ML) has also become an increasingly important component in materials research in the 21st century~\cite{butler2018machine,himanen2019data,schleder2019dft,damewood2023representations}.
One popular ML paradigm in materials science is graph neural networks (GNNs)~\cite{xie2018crystal,schutt2018schnet}, which incorporate both structural and chemical compositional information about materials to make predictions.
Such models have been widely utilized in band gap prediction~\cite{chen2019graph,choudhary2021atomistic,lin2023efficient,moro2023multimodal,li2021graph,tu2024rotnet} as well as in a variety of other materials science tasks~\cite{batzner20223,xie2018hierarchical,chen2021direct,park2020developing,fung2022physically}.

DFT and GNNs are both powerful tools, but also share two common drawbacks.
First, DFT and GNNs both require structural information, which limits their application to materials whose crystal structure is known.
Second, DFT and GNNs are typically relatively ``black box'', in the sense that they do not provide much intuition as to \textit{why} a material has a given band gap.
In contrast, chemists have long had intuition about the band gaps of crystalline materials.
For example, examining the elements present in the chemical formula gives one a sense of the type of bonding that might be expected in the material, which in turn gives some heuristic intuition regarding the presence and size of the band gap (e.g., highly ionic bonds tend to correspond to a large band gap and metallic bonds tend to correspond to zero band gap).
Based on empirical data, relationships between the properties of constituent elements and a material's band gap have long been known for various classes of materials~\cite{manca1961relation,xiuying1992dependence,prokofiev1996periodicity,heng2000prediction,li2012band,korbel2016stability}, and a more recent analysis of a broader dataset also confirmed that the presence or absence of elements affects the distribution of band gaps~\cite{venkatraman2021utility}.
As further evidence of the valuable information contained in the chemical composition, a number of works have demonstrated the use of ML to predict material band gap from just the chemical composition -- these studies have ranged in their type of modeling approach and level of interpretability~\cite{zhuo2018predicting,venkatraman2021utility,goodall2020predicting,ward2016general,wang2021compositionally,prein2023mtencoder,wang2022crabnet,huang2024pretraining}.

Recently, in \citet{ma2023topogivity} and \citet{ma2025chemicalheuristics}, we developed a simple and interpretable ML approach for classifying crystalline materials based on chemical composition alone.
Specifically, for a given material property, this modeling approach features a single parameter for each chemical element.
The classification decision for a given material is then given by sign of the weighted average of the constituent element parameters, where the weighting is with respect to the chemical formula subscripts.
Within the landscape of interpretable models, this represents a particular form of model interpretability -- it is based on the idea of linearly decomposing the decision into contributions from individual elements, and it allows the heuristic interpretation that elements with a greater element parameter have a greater tendency to form materials with the given property.
In general, we would not expect this particular form of model interpretability to work for an arbitrary material property; remarkably, we found that it works surprisingly well for classifying electronic topology and for classifying whether a material is a metal.
Given the known fact that chemical composition does carry substantial information about material band gap as well as the close relationship between the band gap and metallicity modalities, a natural question is whether a similar style of modeling could also be successful for material band gap prediction.

In this paper, we draw on thematic ideas from our previous works on modeling electronic topology and metallicity~\cite{ma2023topogivity,ma2025chemicalheuristics} to develop a simple model for electronic band gap.
From a modeling point of view, a key difference that we need to address in order to adapt the thematic ideas from those works is that the electronic topology and metallicity tasks were treated as discrete binary classification, whereas band gap is not a discrete binary property.
In the existing ML for materials literature, band gap prediction is typically treated as a standard regression problem, often utilizing similar approaches for band gap prediction as for other regression tasks.
However, standard regression tasks typically involve predicting a target variable that can be viewed as a continuous random variable.
In contrast, as we will highlight in this paper, the band gap is actually \textit{not a continous random variable}.
Specifically, in this work, we begin by analyzing the empirical cumulative distribution function of the band gap, which illustrates that the band gap should actually be viewed as a mixed random variable.
Specifically, it can be viewed as a non-negative random variable with a discrete probability mass at exactly 0 eV (electron volts) as well as finite probability density for the positive portion of the support.

As in our previous works on electronic topology and metallicity classification~\cite{ma2023topogivity,ma2025chemicalheuristics}, our band gap model also has a parameter for each element and also involves computing a weighted average of element parameters for each material, with weighting with respect to the chemical formula subscripts.
Rather than looking at the sign of this weighted average, however, here we take the ReLU of this weighted average (i.e., the maximum of this weighted average and zero).
The intention behind this ReLU formulation is to design a model that is tailored for the non-standard statistics of the band gap.
The model provides heuristic chemical interpretability in the sense that elements with greater parameters tend to have greater band gaps.

\section{Problem Setting and Modeling Considerations \label{section_problem_setting_and_modeling_considerations}}

\begin{figure}[!htb]
    \centering
    \includegraphics[width=0.52\textwidth]{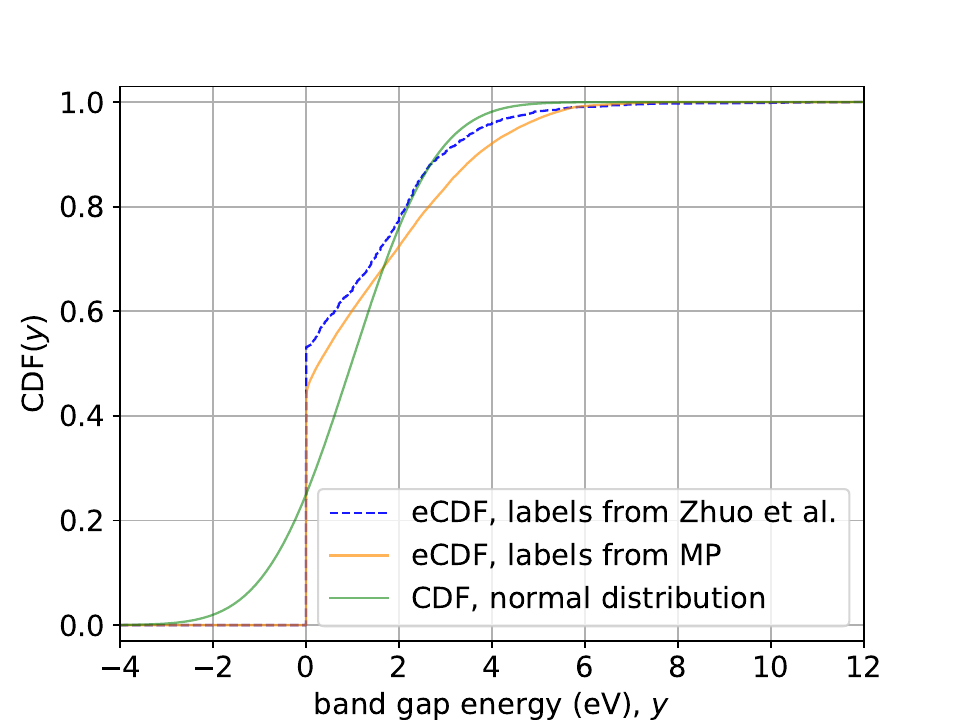}
    \caption{\textbf{The atypical distribution of the band gap.} The empirical cumulative distribution function (eCDF) for electronic band gap is shown for two datasets: a processed version of the Zhuo et al. dataset~\cite{zhuo2018predicting,dunn2020benchmarking} and a dataset based on the Materials Project (MP)~\cite{dunn2020benchmarking,jain2013commentary}.  From its eCDF, we can see that the band gap has a highly non-standard distribution -- if viewed as a random variable, it is neither a purely discrete random variable nor a purely continuous random variable.  For reference, we also show the cumulative distribution function for a normal distribution with mean and variance respectively equal to the sample mean and sample variance of the processed version of the Zhuo et al. dataset.}
    \label{fig_empirical_cdf_data}
\end{figure}

In this work, we are interested in ML models for predicting the electronic band gap of three-dimensional crystalline materials.
We will be in the supervised learning setting, where the material is the input to the model $\hat{\varepsilon}(\cdot)$ and the band gap is the label.
We will take the perspective of viewing the material $M$ and associated band gap label $\varepsilon$ as random variables.

A typical way to quantify the performance of a model $\hat{\varepsilon}(\cdot)$ is to look at the expected loss,
\begin{equation}
    \mathcal{L}\Big(\hat{\varepsilon}(\cdot)\Big) = \mathbb{E}\Big( l \big(\varepsilon, \hat{\varepsilon}(M) \big)\Big), \label{eq_generic_expected_loss}
\end{equation}
where $l(\cdot,\cdot)$ is a loss function (e.g., absolute error or square error)~\cite{hastie2009elements}.
A lower expected loss corresponds to a more accurate model.
Given a suitable set of $N$ pairs of material and band gap label, $\{(M^{(i)}, \varepsilon^{(i)})\}_{i=1}^N$, the expected loss can be estimated using the empirical loss,
\begin{equation}
    \mathcal{L}^{\mathrm{emp}}\Big(\hat{\varepsilon}(\cdot)\Big) = \frac{1}{N}\sum_{i=1}^N l\big(\varepsilon^{(i)}, \hat{\varepsilon}(M^{(i)}) \big). \label{eq_def_empirical_loss}
\end{equation}
In the ML for materials literature, a commonly reported metric is the mean absolute error (MAE), which corresponds to Eq.~\eqref{eq_def_empirical_loss} with the choice $l(a,b)=|a-b|$.

More generally, we can consider the joint cumulative distribution function (CDF)
\begin{equation}
    F_{\mathrm{label, model}}(y_1,y_2) = \mathbb{P}\Big(\varepsilon \leq y_1, \hat{\varepsilon}(M) \leq y_2\Big), \label{eq_joint_CDF_generic_model_and_label}
\end{equation}
where we have chosen to represent the distribution using a CDF here -- as opposed to a probability mass function (PMF) or probability density function (PDF) -- because as we will see, the band gap is neither purely continuous nor purely discrete.
In this work, we will not attempt to characterize the full joint CDF in Eq.~\eqref{eq_joint_CDF_generic_model_and_label}.
Rather, we will focus on the marginal CDF of the labels,
\begin{equation}
    F_{\mathrm{label}}(y) = \mathbb{P}\Big(\varepsilon \leq y\Big),
\end{equation}
and the marginal CDF of the model's predictions,
\begin{equation}
    F_{\mathrm{model}}(y) = \mathbb{P}\Big(\hat{\varepsilon}(M) \leq y\Big).
\end{equation}
For a good model, we expect that $F_{\mathrm{model}}(\cdot)$ should be similar to $F_{\mathrm{label}}(\cdot)$.
Of course, $F_{\mathrm{model}}(\cdot)$ being similar to $F_{\mathrm{label}}(\cdot)$ is \textit{not sufficient} for the model to be good as these are only marginal distributions; even if these marginal distributions are similar one still needs to look at some other way to quantify whether the model actually tends to predict an accurate value, such as by examining the expected loss in Eq.~\eqref{eq_generic_expected_loss} (which carries information about the joint distribution that goes beyond the information contained in the marginal distributions alone).
Nevertheless, in this work, examining these marginal distributions will be valuable both for motivating our model formulation and for characterizing its statistics.

While we do not expect to have analytical expressions for the CDF of the labels ($F_{\mathrm{label}}(\cdot)$) or the CDF of the model predictions ($F_{\mathrm{model}}(\cdot)$), we can approximate them using empirical CDFs~\cite{van2000asymptotic,coles2001introduction,belitz2021evaluation}.
Specifically, given a suitable set of $N$ band gap labels $\{\varepsilon^{(i)} \}_{i=1}^N$, the empirical CDF of the labels is given by
\begin{equation}
    F_{\mathrm{label}}^{\mathrm{emp}}(y) = \frac{1}{N}\sum_{i=1}^N \big[\varepsilon^{(i)} \leq y\big],
\end{equation}
where $\big[\cdot\big]$ denotes the Iverson bracket (takes value 1 when the statement inside is true and zero otherwise).
Similarly, given a suitable set of $N$ model predictions $\{\hat{\varepsilon}(M^{(i)}) \}_{i=1}^N$, the empirical CDF of the model predictions is given by
\begin{equation}
    F_{\mathrm{model}}^{\mathrm{emp}}(y) = \frac{1}{N}\sum_{i=1}^N \big[\hat{\varepsilon}(M^{(i)}) \leq y\big].
\end{equation}

In Fig.~\ref{fig_empirical_cdf_data}, we plot the empirical CDF of the labels for two band gap datasets.
The dashed blue curve corresponds to a processed version of the dataset originally generated by Zhuo et al.~\cite{zhuo2018predicting,dunn2020benchmarking}, which is the dataset that we will use for fitting machine learning models in this paper (note that the processing was done both by MatBench \cite{dunn2020benchmarking} and by us).
This dataset contains 4,603 materials, including both materials that are metals (their 0 eV band gaps were determined by Materials Project~\cite{jain2013commentary} DFT) as well as materials that are not metals (their band gaps were measured experimentally).
We can see that the empirical CDF is zero for all negative values, and then has a large jump at exactly 0 eV, which corresponds to a discrete probability mass at exactly 0 eV band gap.
Subsequently, the empirical CDF increases gradually in the positive portion of the support, and then levels off at 1.
The height of this jump (0.532) indicates the fraction of materials with exactly 0 eV band gap; the fact that it is roughly 0.5 is due to the specific way that \cite{zhuo2018predicting} curated its data.
The solid orange curve corresponds to the matbench-mp-gap dataset~\cite{dunn2020benchmarking,jain2013commentary}, which contains 106,113 materials; these band gaps are originally from DFT calculations from the Materials Project~\cite{jain2013commentary}.
We can see that although the height of the jump at 0 eV is somewhat less than 0.5 for this dataset (specifically, it is 0.435), there is still a substantial proportion of materials with a band gap of exactly 0 eV.
Further details on both datasets are provided in Supplementary Section S1.
For reference, we also plot the CDF for a normal distribution with mean $\mu$ and standard deviation $\sigma$ respectively equal to the sample mean and sample standard deviation of the labels from the processed version of the Zhuo et al. database~\cite{zhuo2018predicting,dunn2020benchmarking}, i.e., $F_{\mathrm{normal}}(y) = \frac{1}{2}\Big(1 + \mathrm{erf}\big( \frac{y-\mu}{\sigma \sqrt{2}} \big) \Big)$, where $\mathrm{erf}(\cdot)$ is the error function (solid green curve).
Since normal distribution corresponds to a continuous random variable, its CDF does not have any jumps.

This makes it clear that the band gap is neither a purely continuous random variable nor a purely discrete random variable.
Rather, it is a mixed random variable.
Specifically, there is a discrete probability mass at exactly 0 eV (corresponding physically to metals) together with a finite probability density for the positive portion of the support (corresponding physically to materials that are not metals).
As such, the band gap is quite different in terms of its statistical properties compared to other properties that are typically treated as standard regression problems; standard regression problems usually involve predicting a purely continuous random variable.

\begin{figure*}[!htb]
    \centering
    \includegraphics[scale=1]{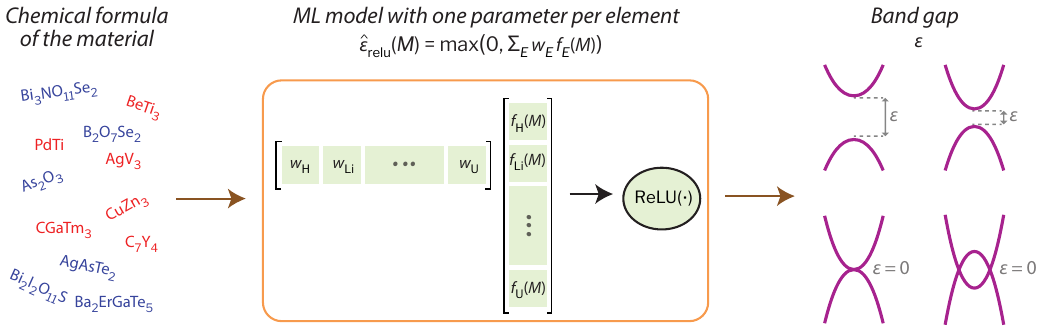}
    \caption{\textbf{Modeling approach based on one parameter for each chemical element.}  The only information that is input into the model is the chemical composition (left panel).  For a given material, the model makes a heuristic prediction of the band gap based on a weighted average of the parameters of the material's constituent elements followed by the ReLU function (center panel). The model is capable of predicting band gap for both non-zero and zero band gap materials (right panel).
    \label{fig_schematic_of_modeling_framework}}
\end{figure*}

\section{Model Description \label{section_model_description}}

For our band gap model, we wish to incorporate the ideas from our previous works on binary classification of materials~\cite{ma2023topogivity,ma2025chemicalheuristics} of having one parameter for each element and computing a weighted average of these parameters with respect to the relative abundances of the elements in the chemical composition.
Additionally, based on our analysis in Section~\ref{section_problem_setting_and_modeling_considerations}, we want our band gap model to be able to predict a substantial fraction of materials to have exactly 0 eV band gap and predict the rest of the materials to have strictly positive band gaps over a continuous distribution.
To achieve these modeling goals, we introduce a model that we will term the ReLU model,
\begin{equation}
    \hat{\varepsilon}_{\mathrm{relu}}(M) = \mathrm{max}\bigg(0, \sum_{E\in\Omega} w_E f_E(M) \bigg) = \mathrm{ReLU}\bigg(\mathbf{w} \cdot \mathbf{f}(M) \bigg). \label{eq_definition_of_epsilon_hat_full_model}
\end{equation}
In the above, $E$ denotes a chemical element and $f_E(M)$ denotes the element fraction for element $E$ in material $M$ (e.g., $f_A(M) = \frac{x}{x+y}$ and $f_B(M) = \frac{y}{x+y}$ for a material with a chemical formula of the form $A_x B_y$).
$f_E(M)=0$ for all elements not present in the material.
$\Omega$ denotes the set of all elements present in the dataset, and $\mathbf{f}(M)$ is the vector of length $|\Omega|$ that contains all of the element fractions for material $M$.
$w_E$ is a learned parameter for each element, $\mathbf{w}$ is the vector containing all of the $w_E$ parameters, and $\hat{\varepsilon}_{\mathrm{relu}}(M)$ is the predicted band gap.
Note that summing $w_E f_E(M)$ over all elements present in the dataset, $\Omega$, is equivalent to just summing $w_E f_E(M)$ over the elements present in the material.
For all materials where the quantity $\mathbf{w} \cdot \mathbf{f}(M) \leq 0$, the model predicts a band gap of exactly 0 eV.
For all other materials, the model predicts the band gap to be $\mathbf{w} \cdot \mathbf{f}(M)$.
As one can see, this means that the model is capable of expressing functions that predict a non-negligible fraction of materials to have exactly 0 eV band gap without needing to have parameter values equal to zero.
Our modeling approach is illustrated schematically in Fig.~\ref{fig_schematic_of_modeling_framework}.

As in our previous works involving binary classification~\cite{ma2023topogivity,ma2025chemicalheuristics}, this model's parameter for each element also allows an element-wise interpretation.
Specifically, the predicted band gap is non-decreasing with respect to an increase in the element parameter value $w_E$, and so we can see that we have the heuristic interpretation that elements with a greater value of $w_E$ tend to form materials with greater band gaps.

The transform ($\mathrm{ReLU}(\cdot)$) that we use to map the element parameter weighted average ($\sum_{E\in\Omega} w_E f_E(M)$) to the predicted band gap ($\hat{\varepsilon}_{\mathrm{relu}}(M)$) has an asymmetric effect.
Specifically, all the negative arguments to this transform map to zero, whereas all the positive arguments are unchanged by this transform.
This means that a very large magnitude negative parameter value would correspond to an element that typically forms materials with 0 eV band gap (i.e., metals), because in the formulation of the model, the term associated with such a parameter value could dominate the rest of the terms in the weighted average.
It would not correspond to an element that tends to form negative band gaps as that is both physically impossible and also not allowed by the modeling framework.
In contrast, if there were hypothetically a large magnitude positive parameter value, then the model would predict that materials with this element tend to have very large gaps, which contrasts with the effect of a large magnitude negative parameter value.
A negative value parameter with arbitrarily large magnitude does not result in physically unreasonable predictions since the model predictions are lower bounded at 0 eV, whereas a positive value parameter with an arbitrarily large magnitude would likely result in physically unreasonable predictions of excessively large band gaps.
In fact, large magnitude negative parameters could actually be necessary for elements that usually form materials with zero band gap, especially for elements which often occur with low element fraction.
As such, there is a strong asymmetry in the interpretation of positive values and negative values for the element parameters in this modeling framework.

To fit the model parameters, we use supervised learning with a labeled dataset containing three-dimensional crystalline materials $M$ and their associated band gaps $\varepsilon$.
Specifically, as previously stated in Section~\ref{section_problem_setting_and_modeling_considerations}, we use a processed version of the Zhuo et al. dataset~\cite{zhuo2018predicting,dunn2020benchmarking}.
Note that since this dataset contains only two, three, and four element materials, the model should be interpreted as applying best to materials with 2-4 elements (in particular, we do not expect it to apply to pure element materials); further details are in Supplementary Section S1.
We train the model using gradient-based optimization by using the Adam algorithm~\cite{kingma2015adam} to minimize the mean square error (MSE).
See Supplementary Section S2 for further details on model fitting.

As a baseline, we also consider a linear model which has the same form as the ReLU model defined in Eq.~\eqref{eq_definition_of_epsilon_hat_full_model} but without the
$\mathrm{ReLU}(\cdot)$.
Specifically, this model takes the form
\begin{equation}
    \hat{\varepsilon}_{\mathrm{linear}}(M) = \sum_{E\in\Omega} w_E f_E(M) = \mathbf{w} \cdot \mathbf{f}(M),
\end{equation}
where $w_E$ is again a learned parameter for each element and $\mathbf{w}$ is again a vector containing all of the $w_E$ parameters.
We optimize the parameters $\{ w_E \}$ for the linear model using ordinary least squares (OLS).
Further details are in Supplementary Section S2.

\section{Empirical Results}

For both the ReLU model as well as the baseline linear model, we evaluate the empirical performance using a 10-fold cross validation approach.
For each split, the dataset is partitioned into train and test sets; we average the test results over all of the splits to obtain the overall test empirical loss and test empirical CDF.
Details on the cross validation procedure and results are provided in Supplementary Section S3.

Table~\ref{table_cross_validation_results} shows the mean and standard deviation of the test MAE from cross validation.
Comparing the test MAE of the ReLU model with the test MAE of the baseline linear model illustrates that the use of the ReLU transform provides significant benefits in terms of reducing error.
(Recall that aside from the use of the $\mathrm{ReLU}(\cdot)$ transform, the formulations of the ReLU model and linear model are the same.)

\begin{table}[!htb]
    \centering
    \begin{tabular}{ l l } 
    \toprule
    \textbf{Model}   \ \ \ \ \ \ \ \ \ \ \ \ & \ \ \ \ \ \ \ \ \ \ \ \  \textbf{Mean Absolute Error (eV)} \\
    \midrule
    Linear model, $\hat{\varepsilon}_{\mathrm{linear}}(\cdot)$ \ \ \ \ \ \ \ \ \ \ \ \ & \ \ \ \ \ \ \ \ \ \ \ \ $0.824 \pm 0.035$  \\ 
    \midrule
    ReLU model, $\hat{\varepsilon}_{\mathrm{relu}}(\cdot)$ \ \ \ \ \ \ \ \ \ \ \ \ & \ \ \ \ \ \ \ \ \ \ \ \ $0.575 \pm 0.036$ \\ 
    \bottomrule
    \end{tabular}
    \caption{\textbf{Empirical error of the model}. Shown are the $\mathrm{mean} \pm \mathrm{standard \ deviation}$ of the test MAE (mean absolute error) from cross validation.  The ReLU model clearly outperforms the baseline linear model.
    }
    \label{table_cross_validation_results}
\end{table}

\begin{figure}[!htb]
    \centering
    \includegraphics[width=0.52\textwidth]{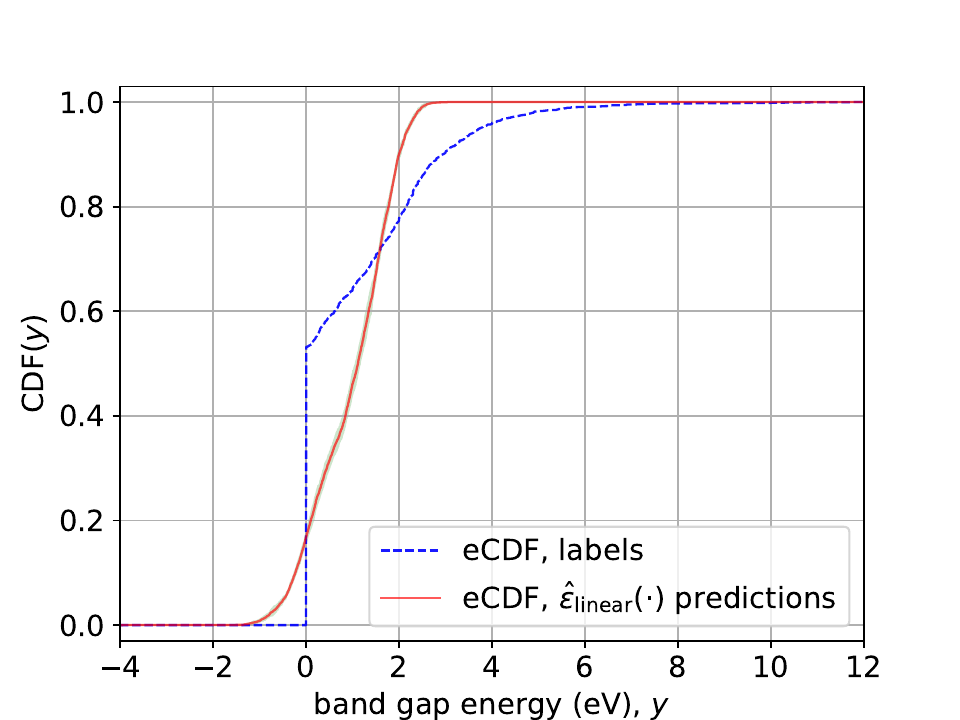}
    \includegraphics[width=0.52\textwidth]{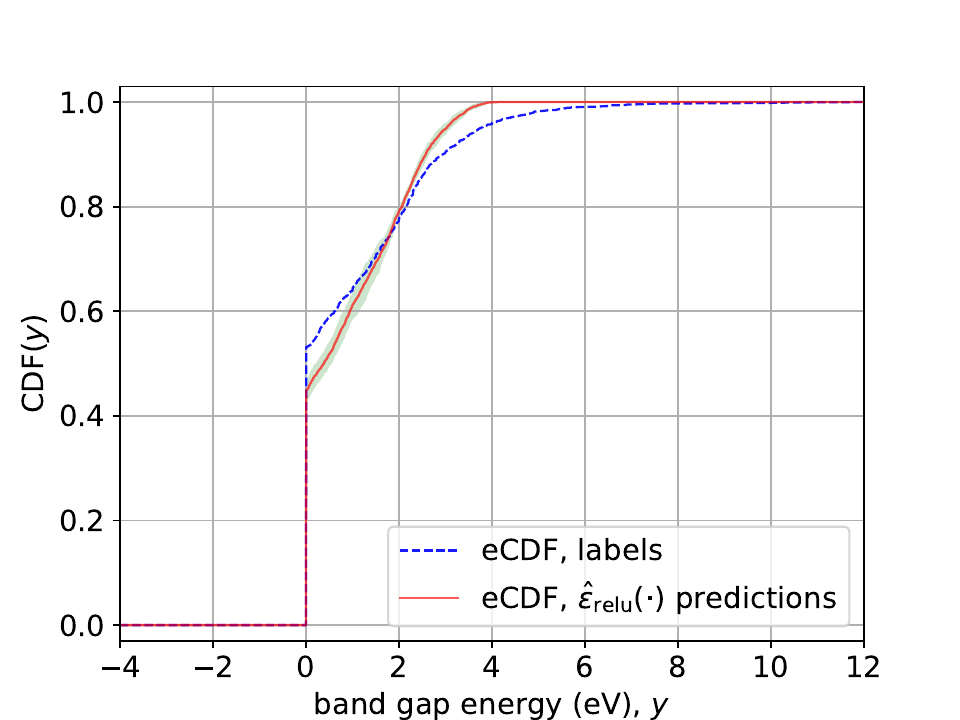}
    \caption{\textbf{Empirical distribution of model predictions}.  The test empirical cumulative distribution function (eCDF) is shown for the linear model's predictions (top) and the ReLU model's predictions (bottom).  The red curves indicate the mean from cross validation and the green shaded region indicates the standard deviation from cross validation.  For comparison, the eCDF of the labels (evaluated using the entire dataset) is also shown as a dashed blue curve in both plots.  We emphasize that from these plots, we can observe that the ReLU model captures the existence of a discrete probability mass at 0 eV (corresponding to metals), whereas the baseline linear model does not.
    \label{fig_empirical_cdfs_of_epsilon_hat}}
\end{figure}

As previously discussed, the band gap has atypical statistics, and our ReLU model was designed to be a model that is suited towards its non-standard distribution.
To further characterize whether the model achieved its design goals, we examine the empirical CDF of the model's predictions.
Fig. \ref{fig_empirical_cdfs_of_epsilon_hat} shows plots of the empirical CDF of the model's predictions for the baseline linear model (top) as well as the ReLU model (bottom).
Also shown for comparison in both plots is the empirical CDF of the labels as evaluated on the full dataset (i.e., the dashed blue curve in these plots is the same as the dashed blue curve in Fig.~\ref{fig_empirical_cdf_data}).
For the baseline linear model, we see that the empirical CDF of the model's predictions does not capture the jump at 0 eV; it also fails to capture the fact that band gap is never negative (it makes a non-negligible amount of negative predictions).
In contrast, for the ReLU model, the empirical CDF of the model's predictions does have a sizable jump at 0 eV; additionally, we can see that this model makes no negative predictions.
Additionally, examining the empirical CDF of the ReLU model's predictions and the empirical CDF of the linear model's predictions in the positive band gap region, we can see that the former is a much better match for the empirical CDF of the labels in the positive portion of the support as well.
Of course, having an empirical CDF of the model's predictions that is similar to the empirical CDF of the labels does not guarantee that the model's predictions are actually accurate, and so the reported empirical loss (in Table~\ref{table_cross_validation_results}) is also essential for characterizing the model's predictive performance. 

\begin{figure*}[!htb]
    \centering
    \includegraphics[scale=0.63]{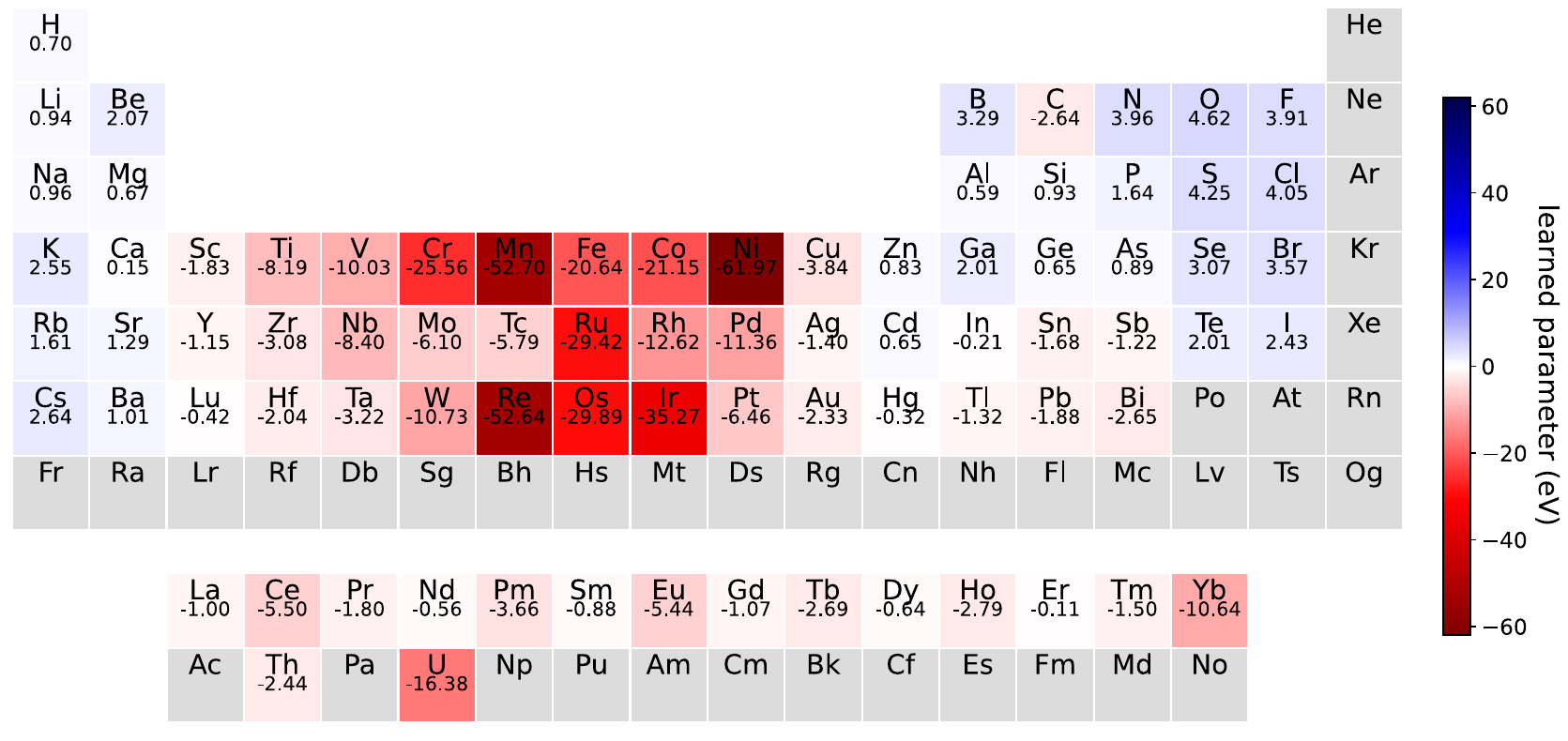}
    \caption{\textbf{Periodic table visualization of the learned parameters}. For each element $E$, its corresponding learned parameter $w_E$ in the ReLU model is indicated numerically and color-coded based on the scale bar (in units of eV).  Elements that are not present in the dataset are displayed in gray.  This visualization illustrates the heuristic chemical interpretability enabled by our simple model for band gap. \label{fig_periodic_table_visualization_of_parameters}}
\end{figure*}

To obtain a final model for visualization, we fit a ReLU model on the entire labeled dataset.
Each learned weight in the ReLU model is shown on the periodic table in Fig.~\ref{fig_periodic_table_visualization_of_parameters}.
We emphasize the simplicity of the model: given a material, the heuristic prediction of the electronic band gap can be achieved by simply looking up the parameter values of its constituent elements in this table and then taking their weighted average -- if the weighted average is positive, then that weighted average is the predicted band gap; otherwise, the predicted band gap is zero.
For example, for the material $\mathrm{BeTi}_\mathrm{3}$ (which has 0 eV ground truth band gap), we can apply the model to get $\hat{\varepsilon}_{\mathrm{relu}}(\mathrm{BeTi}_\mathrm{3}) = \mathrm{ReLU}(\frac{1}{4}w_{\mathrm{Be}} + \frac{3}{4}w_{\mathrm{Ti}}) = \mathrm{ReLU}(\frac{1}{4}(2.07 \ \mathrm{eV}) + \frac{3}{4}(-8.19 \ \mathrm{eV})) = 0 \ \mathrm{eV}$.
As another example, if we apply the model to the material SbSeBr (which has 1.84 eV ground truth band gap), we get $\hat{\varepsilon}_{\mathrm{relu}}(\mathrm{SbSeBr}) = \mathrm{ReLU}(\frac{1}{3}w_{\mathrm{Sb}} + \frac{1}{3}w_{\mathrm{Se}} + \frac{1}{3}w_{\mathrm{Br}}) = \mathrm{ReLU}(\frac{1}{3}(-1.22 \ \mathrm{eV}) + \frac{1}{3}(3.07 \ \mathrm{eV}) + \frac{1}{3}(3.57 \ \mathrm{eV})) = 1.81 \ \mathrm{eV}$.

As discussed in Section~\ref{section_model_description}, the heuristic picture based on the model formulation is that elements with greater parameters typically form materials with greater band gaps (elements with large magnitude negative parameters typically form materials with 0 eV band gap).
As such, Fig.~\ref{fig_periodic_table_visualization_of_parameters} provides a visual illustration of the chemical intuition that the model captures about electronic band gap.
We observe that even though we do not build it into the model \textit{a priori}, it turns out that elements that are close together on the periodic table typically have similar values of learned parameters, which is consistent with human intuition because one expects elements that are close together on the periodic table to be similar.
Additionally, we see that elements on the left and right sides of the periodic table typically have positive values of parameters (whereas elements in the middle of the periodic table typically have negative values of parameters) -- this is also consistent with human intuition, since based on their electronegativities we expect these elements to form ionic compounds, and ionic compounds usually have large band gaps.
The importance of constituent element electronegativities (and in particular, the difference of element electronegativities) for a material's band gap has historically been well appreciated in literature~\cite{manca1961relation,xiuying1992dependence,heng2000prediction,li2012band,korbel2016stability}; this importance has also recently been captured by applying interpretable ML to a narrow class of materials~\cite{yang2023accelerating}.

\section{Discussion and Outlook}

Despite the central importance of the band gap for understanding electronic materials and the substantial research efforts on improving band gap calculation and modeling, band gap prediction still remains an open area of study both on the ab initio methods front and on the ML methods front.
In this work, we developed a new approach for band gap modeling that is based on one parameter for each element.  Our approach for band gap modeling should be viewed as a component of the computational toolkit that is complementary to approaches that are more geared towards predictive accuracy, such as complicated neural network models and expensive ab initio methods such as hybrid functional DFT.
It is important to note that the chemical picture provided by the parameter for each element is only heuristic.
Nevertheless, our model provides advantages of simplicity, interpretability, and essentially instant diagnosis (with the speed advantage being particularly salient when compared against ab initio methods).

In order to extend ideas from our previous works~\cite{ma2023topogivity,ma2025chemicalheuristics} to band gap modeling, we had to look closely at the nature of the band gap, which has atypical statistics in that its distribution is neither purely discrete nor purely continuous.
The fact that the band gap is quite different from standard continuous quantities has sometimes been partially grappled with in the existing literature by taking a two-stage modeling approach where the first model classifies whether the material has a zero or positive band gap and the second model predicts the band gap for those materials that have a positive band gap (e.g., see \cite{isayev2017universal,zhuo2018predicting}).
In our work, it was important for us to have a modeling approach with just a single model, as this enabled us to develop a compact and highly interpretable model with just one parameter for each element.
Another approach sometimes taken in the ML for materials literature is to simply use a dataset that does not contain materials with 0 eV band gap (e.g., see \cite{venkatraman2021utility,moro2023multimodal}).
While that approach also has utility from a methodological development and evaluation point of view, it was important for us to include the 0 eV labels for the goals of this paper because we wanted to design a model that has applicability across the full distribution of band gaps, including the highly atypical part of the distribution at 0 eV.
Moreover, a model that can predict the full range of band gaps has greater utility when given a new material with a completely unknown band gap (in which case it is not known \textit{a priori} whether it is nonzero and so it would be unclear whether one can use a model that only applies to nonzero band gap materials).

Given that many ML works simply treat band gap prediction as a standard regression task, as well as the distinct differences in how we grapple with the non-standard nature of the band gap distribution in comparison with previous approaches such as two-stage modeling, we hope that our framing of band gap as a mixed random variable and our development of a suitable model for its distribution can provide insights of broader relevance.
While the model in this paper was developed with a focus on interpretability and simplicity, our analysis of the highly nonstandard empirical distribution of band gap labels as well as our characterization of the empirical distribution of the model's predictions may also be relevant for machine-learned band gap models that are aimed more at maximizing predictive accuracy.
It could be interesting to draw on relevant ideas from the broader statistics and ML literature on modeling target variables with non-standard distributions (in particular, modeling target variables that take the value zero for a non-negligible portion of the data)~\cite{michaelis2020mixed,lambert1992zero,jorgensen1994fitting,zhou2022tweedie}.

It is also worth pointing out that two materials that both have zero band gap can still actually be fairly different in terms of their band structure -- for example, the sizes of their Fermi surfaces could be very different.
It would be interesting to investigate whether this additional information about the band structure (which is hidden when you only look at the band gap magnitude) could actually be leveraged in some way to improve the ML modeling of the band gap itself within the framework of treating band gap as a mixed random variable.
For example, it might be reasonable to consider a 0 eV material with a large Fermi surface to be a more confident 0 eV label compared to a 0 eV material with a small Fermi surface.

This work represents a contribution to the growing body of work on interpretable ML methods for materials, which is an active area of research across the field of materials science for electronic properties and beyond~\cite{choubisa2023interpretable,moro2023multimodal,venkatraman2021utility,wang2022crabnet,bartel2019new,liu2023materials,ouyang2018sisso,allen2022machine,muckley2023interpretable,chen2022accurate,isayev2017universal,zhang2023physically,ziletti2018insightful,george2020chemist,oviedo2022interpretable,yang2023accelerating,huo2024feature}.
Within the broad landscape of interpretable ML methods for materials, our model represents a particular style of model interpretability, which is based on a weighted average of element parameters.
This style of model interpretability was also featured in our previous works on classification of electronic topology and metallicity~\cite{ma2023topogivity,ma2025chemicalheuristics}.
It would be interesting to adapt this style of interpretable modeling towards other material properties, both to identify new chemical insights and also to characterize the scope of tasks which can be tackled using this type of interpretability.

%----------------------------
%----- ACKNOWLEDGEMENTS -----
%----------------------------
\FloatBarrier

\section*{Acknowledgements}

We thank Thomas Christensen, Liang Fu, Charlotte Loh, Rumen Dangovski, Ian Hammond, David Dai, Julian Powers, Andr\'e Grossi Fonseca, Ali Ghorashi, and Lindley Winslow for helpful discussions and comments.
A.M. would like to acknowledge support for his PhD from the National Science Foundation Graduate Research Fellowship under Grant No. 1745302.
O.D. would like to thank the Hertz Foundation for their support of his Ph.D.
This material is based upon work sponsored in part by the U.S. Army DEVCOM ARL Army Research Office through the MIT Institute for Soldier Nanotechnologies under Cooperative Agreement number W911NF-23-2-0121.
It is also supported in part by the Air Force Office of Scientific Research under the award number FA9550-21-1-0317, as well as in part by the National Science Foundation under Cooperative Agreement PHY-2019786 (The NSF AI Institute for Artificial Intelligence and Fundamental Interactions, \url{http://iaifi.org/}).

\section*{Code availability}
The code used to produce the results in this paper will be made available in a public repository.

\def\bibsection{\section*{\refname}}
\bibliographystyle{apsrev4-2-longbib}
\bibliography{refs}

\end{document}